\documentclass[%
 reprint,
nofootinbib,
 amsmath,amssymb,
 aps,
]{revtex4-2}

\usepackage{graphicx}
\usepackage{dcolumn}
\usepackage{bm,xcolor}
\usepackage{hyperref}
\usepackage{physics}
\usepackage{comment}
\hypersetup{colorlinks=true, allcolors=blue}

\begin{document}

\preprint{APS/123-QED}

\title{Checking volume-law entropy with Hubeny-Rangamani-Takayanagi surfaces}

\author{Tyler Guglielmo}
 \affiliation{Department of Physics, Theory Group, The University of Texas at Austin, Austin, Texas 78712, USA} 
 
\author{Phuc Nguyen}%
\affiliation{Department of Physics and Astronomy, Lehman College, City University of New York,
250 Bedford Park Boulevard West, Bronx, New York 10468, USA}
\affiliation{Department of Mathematics and Haifa Research Center for Theoretical Physics and Astrophysics,
University of Haifa,
Haifa 3498838, Israel}

\begin{abstract}
We check formally that the Hubeny-Rangamani-Takayanagi prescription for holographic entanglement entropy -- when applied to a static black brane spacetime and to a wide class of subregions that do not lie on a constant time slice -- gives rise to volume-law entropy in the limit of large subregion. By volume-law entropy, we mean that the entanglement entropy scales with the volume of the projection of the boundary subregion onto a static time slice with respect to the boundary thermal state (with the same coefficient as in the volume law on the static time slice). Our result applies to subregions that have reflection symmetry as well as strips, and we also present field-theoretic arguments in support of our holographic findings.
\end{abstract}

\begin{flushright}
	\texttt{UTTG 29-2022}
\end{flushright}

\maketitle

\tableofcontents

\section{Introduction}
The entanglement entropy of a subregion of a field theory in a thermal state has an extensive component that scales with the volume of the subregion, when the subregion is large. In holography, previous work by Liu-Mezei \cite{Liu:2012eea, Liu:2013una} has shown that the Hubeny-Rangamani-Takayanagi (HRT) surface \cite{Ryu:2006bv,Ryu:2006ef,Hubeny:2007xt} is consistent with this volume scaling in static situations. See also \cite{Liu:2013iza,Liu:2013qca, Mezei:2016zxg,Mezei:2018jco} for related techniques. In time-dependent situations, it is widely expected that the HRT surface continues to be consistent with the volume law, at least for an appropriate class of nonthermal states. In this paper, we take the first steps toward generalizing the Liu-Mezei technique to a particular kind of nonstatic situation: we will consider entangling subregions that break staticity in thermal states (and not states with general time dependence).

The rest of the paper is organized as follows. In Sec. \ref{Sec:FieldTheory}, we describe our intuition from the field theory side. In Sec. \ref{Sec:Strips}, we check the volume law for strips lying on a boosted time slice on the boundary of a black brane spacetime. In Sec. \ref{Sec:General}, we present a more general argument for any subregion that has reflection symmetry. Finally, in Sec. \ref{Sec:Conclusion}, we conclude. Some of the more technical results are relegated to the Appendixes.

\section{Expectation from the field theory side}\label{Sec:FieldTheory}
\begin{figure}
\includegraphics[width=6cm]{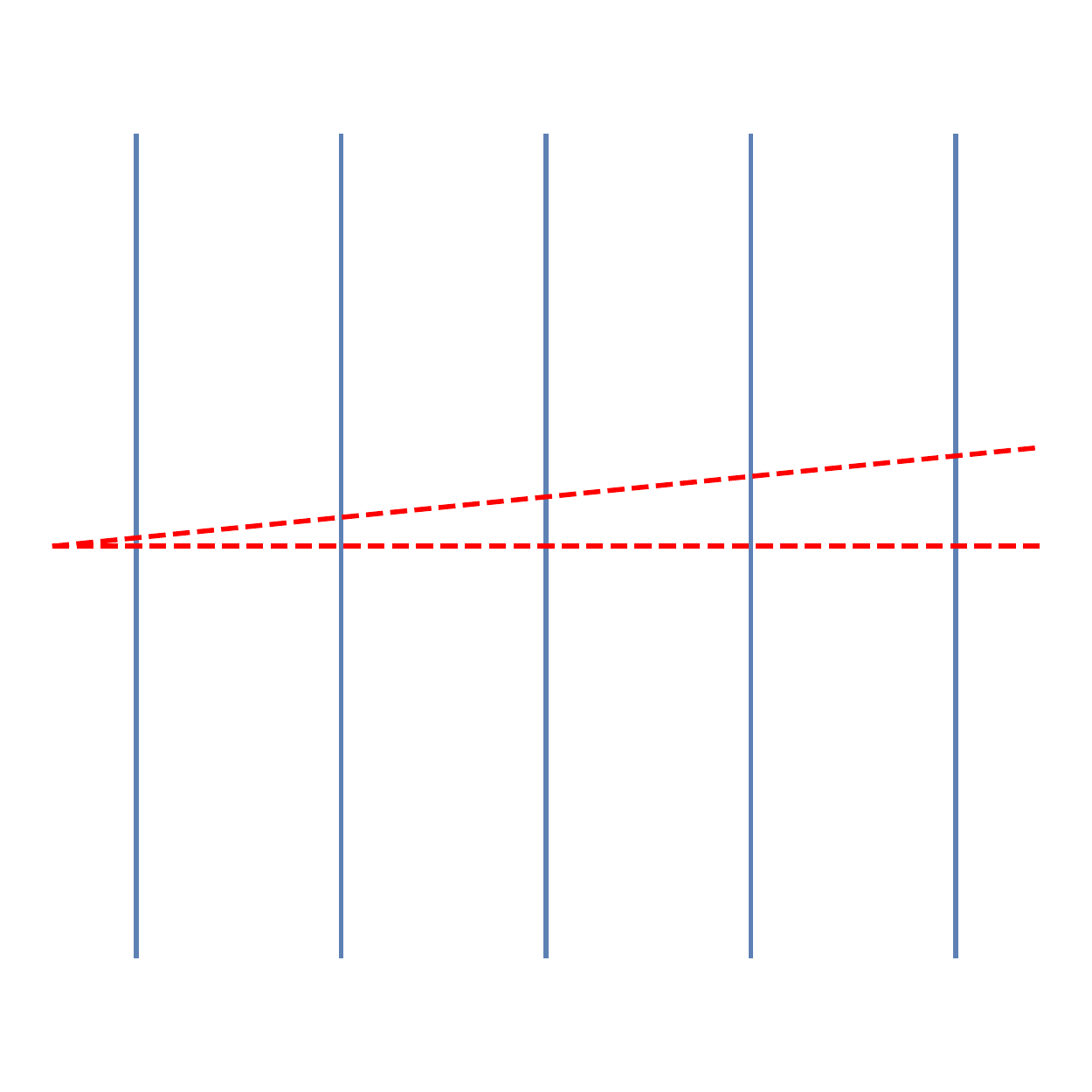}
\caption{The worldlines of the qubits are in blue. The unboosted and boosted subregions are in red.}
\label{fig:QubitWorldlines}
\end{figure}

Intuitively, we expect that the entanglement entropy of the boosted thermal state is proportional to the volume of the projection of the subregion onto a ``static time slice'' (i.e., a slice on which the stress-energy tensor of the field theory is diagonal). To see this, we can consider a system of qubits at equal separation in space, each of which is in a maximally mixed state (so, in particular, the qubits are not entangled among them). Furthermore, let us suppose that the qubits are stationary, so their worldline is perpendicular to the $t=0$ slice, as depicted in Fig. \ref{fig:QubitWorldlines}.

Now consider two subregions (depicted in red in Fig. \ref{fig:QubitWorldlines}), one of which is on the unboosted time slice and the other one is on a boosted time slice, such that the projection of the boosted subregion onto the unboosted time slice is the unboosted subregion. Simply by counting the number of qubits contained in the subregions, we see that the entanglement entropy of the two subregions are equal to each other. In other words, the entanglement entropy of a boosted subregion scales with the volume of the projection onto a static time slice, with the same coefficient as in volume-law entropy on the static time slice.

Next, we back up this intuition with a few simple computations. Consider a field theory in the thermal state $\rho = \frac{e^{-\beta H}}{Z}$ with $Z = \mathrm{Tr}{(e^{-\beta H})}$. This is the state on the $t=0$ slice. We now boost the state in the $z$ direction by conjugating $\rho$ with the unitary operator implement the boost. Using the Poincar\'{e} algebra, we find the new state on the $t=0$ slice to be $\rho_{boosted} = \frac{e^{-\beta H_{boosted}}}{Z}$ with $H_{boosted} = \gamma H - v \gamma P_{z}$, where $v$ is the boost velocity, $\gamma$ is the dilation factor $\gamma = 1 / \sqrt{1 - v^2}$, and $P_{z}$ is the translation generator along the $z$ direction.

It will be convenient to rewrite the boosted state in the Gibbs form,
\begin{equation}
    \rho_{boosted} = \frac{e^{-\beta_{boosted} (H - vP_{z})}}{Z} 
\end{equation}
with $\beta_{boosted} = \gamma \beta$ the ``boosted temperature,'' and with $v P_{z}$ playing the role of a work term similar to the product of the chemical potential and the number of particles. Thus, the state on the $t=0$ slice is a grand canonical ensemble with the unconventional feature that the temperature is $v$ dependent.

Computing the von Neumann entropy $S_{vN} \equiv -\mathrm{Tr}{(\rho \ln \rho)}$ of the state, we find
\begin{equation}\label{SvN}
    S_{vN} = \beta_{boosted} \langle H - vP_{z} \rangle_{boosted} + \ln Z
\end{equation}
and recognize the formula for the Gibbs entropy.

\subsection{Free massless scalar field}
We now specialize to the free massless scalar field in $d+1$ dimensions, and compute the first term of (\ref{SvN}). The expectation value $\langle H - vP_{z} \rangle_{boosted}$ is
\begin{equation}
    \langle H - vP_{z} \rangle_{boosted} = V \int \frac{d^{d}k}{(2\pi)^d} \frac{|\textbf{k}| - vk_{z}}{e^{\beta_{boosted}(|\textbf{k}| - vk_{z})} - 1},
\end{equation}
where $V$ is the proper volume on the $t=0$ slice. Using spherical coordinates in momentum space,
\begin{equation}
    \langle H - vP_{z} \rangle_{boosted} = \frac{V}{(2\pi)^d} \int d\Omega_{d-1} \int dk k^{d-1} \frac{k(1 - v\cos{\theta})}{e^{\beta_{boosted} k (1-v\cos{\theta})}-1},
\end{equation}
where $\theta$ is the polar angle. By doing the change of variable $q = \beta_{boosted} k (1- v\cos{\theta})$, one can show that the integral above is $v$ independent. So we find
\begin{equation}
    \langle H - v P_{z} \rangle_{boosted} = \langle H \rangle_{unboosted}.
\end{equation}

The second term of (\ref{SvN}) can be computed in a similar manner, and yields very similar results. So, the entropy is $S_{thermal} \sim \beta_{boosted} \langle H - vP_{z} \rangle_{boosted}$, which is also $\beta_{boosted} \langle H \rangle_{unboosted}$. So the $v$ dependence in the thermal entropy is through a Lorentz boost factor contained in the boosted temperature. That boost factor can be combined with the proper volume $V$ on the $t=0$ slice to give the projected volume onto the time slice on which the stress tensor is diagonal. Thus, we have checked the intuition described in Fig. \ref{fig:QubitWorldlines} in the setting of a free scalar field.

\subsection{Conformal perfect fluid}
Next, instead of a massless free scalar field, let us specialize Eq. (\ref{SvN}) to the case of a conformal fluid in $d+1$ dimensions. The stress-energy tensor of such a fluid takes the form
\begin{equation}\label{ConformalFluidStress}
    \langle T^{\mu\nu} \rangle = \alpha T^{d+1} (\eta^{\mu\nu} + (d+1)u^{\mu}u^{\nu}) 
\end{equation}
for some constant $\alpha$, and where $u^{\mu}$ is the four velocity of the fluid. This stress tensor is that of a conformal field theory (CFT) dual to a boosted Schwarzschild-anti-de Sitter (AdS) black hole, so we would expect that lessons learned here might carry over to other holographic scenarios.  In addition, the Hamiltonian and translation generator are given in terms of the stress-energy tensor by the standard formulas
\begin{align}
    H &= \int T_{tt} ~d^{d}x,\\
    P_{z} &= \int T_{zt} ~d^{d}x.
\end{align}
We now compute the $\langle H - vP_{z} \rangle_{boosted}$ in (\ref{SvN}). We have
\begin{equation}
    \langle H - vP_{z} \rangle_{boosted} = \int \left( \langle T_{tt} \rangle_{boosted} - v \langle T_{zt} \rangle_{boosted} \right) d^{d}x,
\end{equation}
where $\langle T_{\mu\nu} \rangle_{boosted}$ can be obtained from (\ref{ConformalFluidStress}) with the four velocity taken to be $u^{t} = \gamma$, $u^{z} = -\gamma v$ (note the negative sign) and all other spatial components vanishing,
\begin{align}
    \langle T_{tt} \rangle_{boosted} &= \alpha T^{d+1} (-1 + (d+1)\gamma^{2}),\\
    \langle T_{zt} \rangle_{boosted} &= \alpha T^{d+1} (d+1) \gamma^{2} v.
\end{align}
We then find
\begin{equation}
    \langle H - vP_{z} \rangle_{boosted} = \alpha T^{d+1} d \int d^{d}x.
\end{equation}
Now let us compare the answer above with $\langle H \rangle_{unboosted}$. We have
\begin{equation}
    \langle H \rangle_{unboosted} = \int \langle T_{tt} \rangle_{unboosted} ~d^{d}x,
\end{equation}
where $\langle T_{tt} \rangle_{unboosted}$ is given by (\ref{ConformalFluidStress}) with $u^{t} = 1$ (and all other components vanishing),
\begin{equation}
    \langle T_{tt} \rangle_{unboosted} = \alpha T^{d+1} d.
\end{equation}
We therefore find
\begin{equation}
    \langle H - vP_{z} \rangle_{boosted} = \langle H \rangle_{unboosted},
\end{equation}
just like in the free scalar field case. Beyond this point, the steps are the same as for the free scalar field case: the von Neumann entropy is $\sim \beta_{boosted} \langle H - vP_{z} \rangle_{boosted} = \beta_{boosted} \langle H \rangle_{unboosted}$. The $\gamma$ factor from the boosted inverse temperature can be combined with the volume factor, yielding the projected volume.

\subsection{Projected volume versus other volumes}
\emph{A priori}, one might naively think that the entropy of a subregion lying on a boosted time slice (in the thermal state) should scale with the proper volume of the subregion on that time slice. To see why, let us consider a subregion $A$ lying on an unboosted time slice, and the unitary operator $U$ which implements the boost. $U$ is the exponential of a local integral of the stress-energy tensor (i.e., the boost generator), so it might seem that $U$ is the tensor product of local unitaries,
\begin{equation}
    U \stackrel{?}{=} U_{A} \otimes U_{\bar{A}},
\end{equation}
where $U_{A}$ is the exponential of the integral over the subregion $A$, and $U_{\bar{A}}$ is the exponential of the integral over the complement $\bar{A}$. If the equation above is true, then we would conclude that $U$ has no effect on the entanglement entropy, and thus the volume law is still with respect to the proper volume on the (now boosted) slice.

However, we know that $U$ does not have the local unitary form as written above (see \cite{Kabat:2020oic, Kabat:2021akg} for recent studies of the effects on modular Hamiltonians of the failure of such unitaries to be local). An intuitive way to understand why $U$ does not factorize is by noting that the stress-energy tensor moves local operators around, and as such the ``factorization'' is a mixed combination of $A$ and $\bar{A}$. Thus, we think that the scaling of the entropy with the projected volume rather than the proper volume is a consequence of a nonlocal boost unitary.

\subsection{How universal is the projected volume scaling?}
It is easy enough to find counterexamples to the projected volume statement if we consider states other than a thermal state. For example, a quenched state -- where entropy evolves in time -- is easily seen to be a counterexample.  Hence, we believe the projected volume law should only hold in a thermal state while holding true for subregions of arbitrary shape.  However, this projected volume law should receive corrections after moving away from a thermal state.

It might be surprising that such a universal behavior can be true, given that entanglement entropy for a CFT on the torus is nonuniversal and depends on the operator contents of the theory \cite{Calabrese:2009qy}. However, we note that the projected volume law is a statement only for very large subregions (larger than the thermal scale). So, effectively, we are in a high-temperature regime where it is conceivable that universal statements can be made.

\subsection{Other comments}
We also note the argument in \cite{Casini:2015zua}, where the authors used strong subadditivity to argue in a special case that the entanglement entropy on a null strip is equal to the entanglement entropy of its projection.

\section{A simple holographic example: strips on a boosted time slice}\label{Sec:Strips}
We now move on to discuss holography. In this setting, we can also expect the scaling of the entropy with the projected volume to come out, basically because of the coordinate singularity at the horizon of a black hole spacetime.

We note that what we are after is a finite contribution to the entropy, as illustrated from the field theory examples of the previous section. As such, we do not need to use an explicit renormalization scheme to deal with the UV-divergent part (i.e. the vacuum part) of the entanglement entropy. Holographically, we impose a fixed near-boundary cutoff at some small value of the Fefferman-Graham coordinate $z = \epsilon$, and send the size of the boundary subregion to infinity while keeping $\epsilon$ fixed.

\subsection{Boosted Black Brane}
We first consider a simple example: a strip boundary subregion in a $3+1$-dimensional boosted black brane (i.e., the strip lies on a boosted time slice on the boundary). The metric of the Schwarzschild black brane is
\begin{equation}
    ds^{2} = \frac{L^2}{z^2} \left[ -f(z) dt^{2} + \frac{dz^2}{f(z)} + dx^{2} +dy^{2} \right],
\end{equation}
with $f(z) = 1 - \frac{z^3}{z_{h}^{3}}$, and $z_{h}$ is the $z$ coordinate of the horizon. Throughout this paper, we will set $z_{h} = 1$ for convenience. We could work with boosted coordinates, but it will be simpler to work with the static coordinates above. The strip is delimited by the two lines at $x= \pm R$, $t = \pm t_{0}$ for some half-width $R$ and time $t_{0}$. We parametrize the HRT surface as $t(z)$, $x(z)$. The induced metric on a codimension-2 spacelike surface anchored at the strip is
\begin{equation}
    ds^{2} = \frac{L^2}{z^2} \left[ \left( \frac{1}{f} - ft'^{2} + x'^{2} \right)dz^{2} + dy^{2} \right],
\end{equation}
and the area functional $\mathcal{A}$ of the surface (evaluated over a segment of length $\Delta y$ along the $y$ direction) is
\begin{align}
    \mathcal{A} &= 2 \Delta y \int_{\delta}^{z_m} dz \frac{L^2}{z^2} \sqrt{Q},\\
    Q &= \frac{1}{f} - ft'^{2} + x'^{2},
\end{align}
where $\delta$ is a near-boundary cutoff, and $z_{m}$ is the $z$ coordinate of the ``tip'' of the HRT surface. The Lagrangian above is independent of $t$ and of $x$, hence there are two conserved momenta:
\begin{align}
    p_{t} &= \frac{\partial \mathcal{L}}{\partial t'} = -\left( \frac{L}{z} \right)^{2} \frac{f t'}{\sqrt{Q}},\\
    p_{x} &= \frac{\partial \mathcal{L}}{\partial x'} = \left( \frac{L^2}{z^2} \right) \frac{x'}{\sqrt{Q}}.
\end{align}
We can solve algebraically for $x'$ and $t'$ as functions of $z$,
\begin{align}
    x' &=  \frac{p_x}{\sqrt{p_{t}^{2} + f \left( \frac{L^4}{z^4} - p_{x}^{2} \right)}},\\
    t' &= - \frac{p_t}{f} \frac{1}{\sqrt{p_{t}^{2} + f \left( \frac{L^4}{z^4} - p_{x}^{2} \right)}},
\end{align}
where we have assumed $x' < 0$ and $t' < 0$. These assumptions should hold true over half the HRT surface, and on the other half we have $x' > 0$ and $t' >0$. Note that these assumptions imply $p_{x} < 0$ and $p_{t} > 0$.
We can then integrate the two equations above to get the HRT surface,
\begin{align}\label{xstrip}
    x(z) &= R + \int_{0}^{z} dz' \frac{p_x}{\sqrt{p_{t}^{2} + f(z') \left( \frac{L^4}{z'^4} - p_{x}^{2} \right)}},\\
	\label{tstrip}
    t(z) &= t_{0} - \int_{0}^{z} dz' \frac{p_t}{f(z')} \frac{1}{\sqrt{p_{t}^{2} + f(z') \left( \frac{L^4}{z'^4} - p_{x}^{2} \right)}}.
\end{align}
The two constants $R$ and $t_{0}$ are determined by the momenta $p_{x}$ and $p_{t}$. To see that, we evaluate the two functions above at $z=z_{m}$ and use the fact that $x(z_m) = t(z_m) = 0$. To see that the $x$ and $t$ coordinates of the tip of the HRT surface are zero, we note that both the boundary subregion and the bulk metric are symmetric under the double reflection $x \rightarrow -x$ and $t \rightarrow -t$. Therefore, the tip of the HRT surface must be preserved under that reflection, hence it is located at $x=t=0$. We then obtain
\begin{align}
    R &= - \int_{0}^{z_m} dz' \frac{p_x}{\sqrt{p_{t}^{2} + f(z') \left( \frac{L^4}{z'^4} - p_{x}^{2} \right)}},\\
    t_{0} &= \int_{0}^{z_m} dz' \frac{p_t}{f(z')} \frac{1}{\sqrt{p_{t}^{2} + f(z') \left( \frac{L^4}{z'^4} - p_{x}^{2} \right)}}.
\end{align}
In the two equations above, $z_{m}$ itself is of course determined by $p_{t}$ and $p_{x}$, via the condition that $x'$ blows up at $z=z_{m}$. Explicitly, this condition is
\begin{equation}
    p_{t}^{2} + f(z_m) \left( \frac{L^4}{z_{m}^4} - p_{x}^{2} \right) = 0.
\end{equation}
Let us solve for $p_{t}$ as a function of $z_{m}$ and $p_{x}$ from the above, and plug into the two previous equations. Note that the above forces a lower bound on $|p_{x}|$,
\begin{equation}\label{pxlowerbound}
|p_{x}| \geq \frac{L^2}{z_{m}^{2}}.    
\end{equation}
We will call the right-hand side above the negative of the critical value of $p_{x}$ and denote it by $|p_{x,crit}|$. We then obtain
\begin{equation}\label{Rintegral}
    R = - \int_{0}^{z_m} dz' \frac{p_x}{\sqrt{H(z')}},
\end{equation}
\begin{equation}\label{t0integral}
    t_{0} = \int_{0}^{z_m} \frac{dz'}{f(z')} \frac{\sqrt{-f(z_{m}) \left( \frac{L^4}{z_{m}^{4}} - p_{x}^{2} \right)}}{\sqrt{H(z')}},
\end{equation}
with
\begin{equation}
H(z) \equiv f(z) \left( \frac{L^4}{z^4} - p_{x}^{2} \right) - f(z_{m}) \left( \frac{L^4}{z_{m}^{4}} - p_{x}^{2} \right).
\end{equation}
The two equations above tell us how $R$ and $t_{0}$ are related to $p_{x}$ and $z_{m}$. The area functional becomes:
\begin{equation}\label{Area}
    \mathcal{A} = 2 \Delta y \int_{\delta}^{z_m} dz' \frac{L^4}{z'^4} \frac{1}{\sqrt{H(z')}}.
\end{equation}
Next, we take the large-$R$ limit, with the ratio $R/t_{0}$ kept fixed. We want to keep this latter ratio fixed so that the strip remains on the same boosted time slice as we make the half-width larger and larger. We expect that $z_{m} \approx z_{h} = 1$ in the large-$R$ limit (indeed it is well known that the horizon acts as a barrier to extremal surfaces \cite{Hubeny:2012ry,Engelhardt:2013tra}). So, for the rest of this section, $z_{m}$ will be taken to be close to the horizon. We also need to find the approximate value of $p_{x}$. To do so, we need to think about how the function $H(z)$ looks like.

\begin{figure}
\includegraphics[width=8cm]{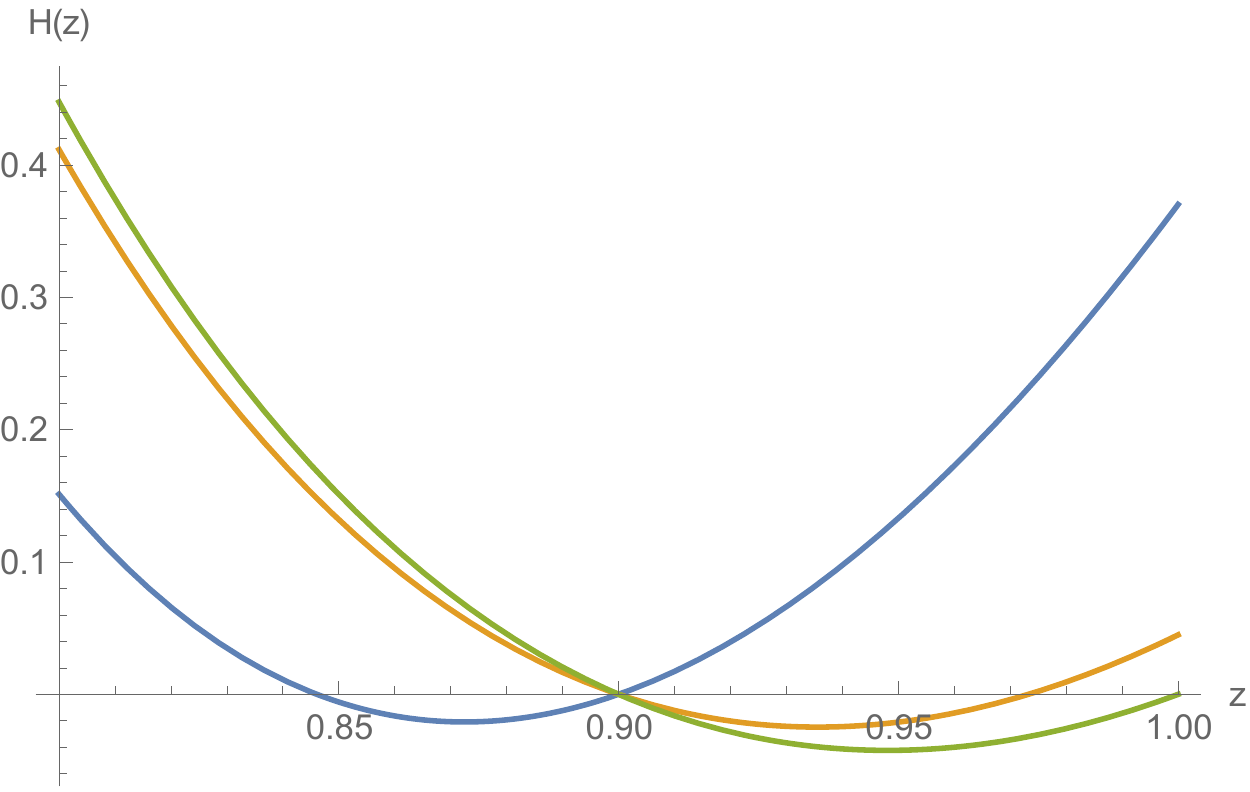}
\caption{Plot of $H(z)$ versus $z$ for three different choices of parameters. For all three curves, we set $L=1$, $z_{m}=0.9$, and $z_{h}=1$. The values of $p_{x}$ differ from curve to curve: $p_{x}=-1.7$ (blue), $p_{x} = -1.3$ (yellow), and $p_{x}=(L/z_{m})^{2} \approx -1.234$ (green). The blue curve is in the unphysical parameter region, since $z_{m}$ is the larger of the two zeros. The yellow and green curves are in the physical parameter region. At the critical value of $p_{x}$, the two roots of $H(z)$ are $z_{m}$ and $z_{h}$, as can be seen by inspecting the green curve.}
\label{fig:Hplot}
\end{figure}

For generic values of $z_{m}$, $z_{h}$, and $p_{x}$, the function $H(z)$ has two zeros, one of which is at $z = z_{m}$, and a minimum between the two zeros. Let $z = z_{M}$ denote the location of this minimum. In addition to the lower bound (\ref{pxlowerbound}), $p_{x}$ is also constrained by the fact that we should only consider values of $p_{x}$ for which $z_{m}$ is the smaller of the two zeros. Indeed, in the other case, we have that $H(z)$ is negative when $z$ approaches $z_{m}$. But $H(z)$ occurs under a square root in the $R$ integral (\ref{Rintegral}), so the integrand is not real near the upper limit of integration (and the integral has to be real). We plot $H(z)$ for a few representative values of the parameters in Fig. \ref{fig:Hplot}.\\

From the plot of $H(z)$ and Eq. (\ref{Rintegral}), we see that the large-$R$ regime corresponds to the regime where $z_{M}$ is close to $z_{m}$. In this regime, we can check this divergence of $R$ by replacing the $H(z)$ in (\ref{Rintegral}) with its expansion to second order around the minimum,
\begin{equation}\label{quadraticH}
    H(z) = H(z_{M}) + H_{2} (z-z_{M})^{2} + \dots,
\end{equation}
where
\begin{equation}
    H_{2} = \frac{1}{2}f''(z_{M}) \left( \frac{L^4}{z_{M}^{4}} - p_{x}^{2} \right) - \frac{4L^4}{z_{M}^{5}} f'(z_{M}) + \frac{10L^{4}}{z_{M}^{6}} f(z_{M}).
\end{equation}
The precise value of $H_{2}$ will not be important for our purposes. The $R$ integral then takes the form
\begin{equation}
    R \approx -\int_{z_{r} - \delta}^{z_{r}} dz' \frac{p_{x}}{\sqrt{H(z_{M}) + H_{2}(z' - z_{M})^{2}}},
\end{equation}
where $z_{r}$ is the zero of the quadratic expression (\ref{quadraticH}), which is smaller than $z_{M}$, and $\delta$ is some small number.\footnote{We note that although $\delta$ is small -- it is not an infinitesimal quantity. We are expanding in the regime where $z_M - z_r$ is small.} Evaluating the integral, we find
\begin{equation}
    R \approx \frac{p_{x}}{2\sqrt{H_{2}}} \log{|H(z_{M})|}.
\end{equation}
So, $R$ diverges when $H(z_{M}) = 0$, which is to say that the zero at $z=z_{m}$ is also a minimum of $H(z)$. To find out which value of $p_{x}$ corresponds to that, we solve for $p_{x}$ from the equation $H'(z_{m}) = 0$, and find
\begin{equation}
    p_{x} = p_{x,crit}.
\end{equation}
So, $p_{x}$ has to be the critical value for $R$ to diverge.

We now replace the $p_{x}$ in the $R$ integral (\ref{Rintegral}) by the critical value
\begin{equation}
    R \approx \int_{0}^{z_{m}} dz' \frac{|p_{x,crit}|} {\sqrt{H(z')}}.
\end{equation}
Also, we can approximate the area functional (\ref{Area}) by setting some of the $z'$ dependence in the integrand to $z_{m}$ (which is near the horizon),
\begin{equation}
    \mathcal{A} \approx 2 \Delta y \int_{\delta}^{z_{m}} dz' \frac{L^{2}}{z_{m}^{2}} \frac{|p_{x,crit}|} {\sqrt{H(z')}}.
\end{equation}
Comparing the two previous equations, and dividing by $4 G_{N}$, we see that the entanglement entropy $S$ is
\begin{equation}
    S \approx s 2 R \Delta y,
\end{equation}
with the entropy density $s$ given by
\begin{equation}
    s = \frac{L^2}{4 G_{N} z_{h}^{2}}
\end{equation}
(where we restored $z_{h}$ for clarity). Thus, we have obtained the statement of volume-law entropy, with the volume given by $2 R \Delta y$. We note that this is the volume of the projection of the strip onto a static time slice, which is independent of the boost angle (or the ratio $R/t_{0}$) and not the proper volume of the strip. We also note that, in the analysis above, we did not need to find out how the ratio $R/t_{0}$ depends on $z_{m}$ and $p_{x}$: for all values of the boost angle, we have that $z_{m} \approx z_{h}$ and $p_{x} \approx p_{x,crit}$ in the large-$R$ limit. The exact boost angle of the strip will depend on how close $z_{m}$ is to $z_{h}$, in comparison with how close $p_{x}$ is to $p_{x,crit}$.

\subsection{The special case of the BTZ black hole}
Finally, we discuss the special case of the Ba\~{n}ados-Teitelboim-Zanelli (BTZ) black hole, where boundary subregions are simply intervals. In this case, a more explicit treatment can be given when compared to the higher-dimensional cases, since geodesics anchored at the two end points of a boosted boundary interval can be found explicitly.

Our strategy to find these geodesics will be to map from the Poincar\'{e} patch, where geodesics are easy to write down, to BTZ by a coordinate transformation. The three-dimensional Poincar\'{e} patch metric is
\begin{equation}
    ds^{2} = \frac{L^2}{z^2} (-dt^{2} + dx^{2} + dz^{2}).
\end{equation}
The geodesic connecting the point $(t=0,x=0)$ on the boundary of AdS to the point $(t=t_{0},x=x_{0})$ on the boundary of AdS can be given in parametric form by
\begin{equation}\label{zofu}
    z(u) = \sqrt{u(1-u)(-t_{0}^{2} + x_{0}^{2})},
\end{equation}
\begin{equation}\label{tofu}
    t(u) = ut_{0},
\end{equation}
\begin{equation}\label{xofu}
    x(u) = ux_{0},
\end{equation}
where $u$ is a parameter ranging from $0$ to $1$. To obtain the expressions above, we can start by a geodesic on the $t=0$ time slice in the Poincar\'{e} patch and boost it (along the boundary direction $x$). The result is guaranteed to be a geodesic because boosting along $x$ is an isometry of the spacetime. By eliminating the parameter $u$, we can also describe the geodesics by the two functions $t(x)$ and $z(x)$.

\begin{figure}
\includegraphics[width=8cm]{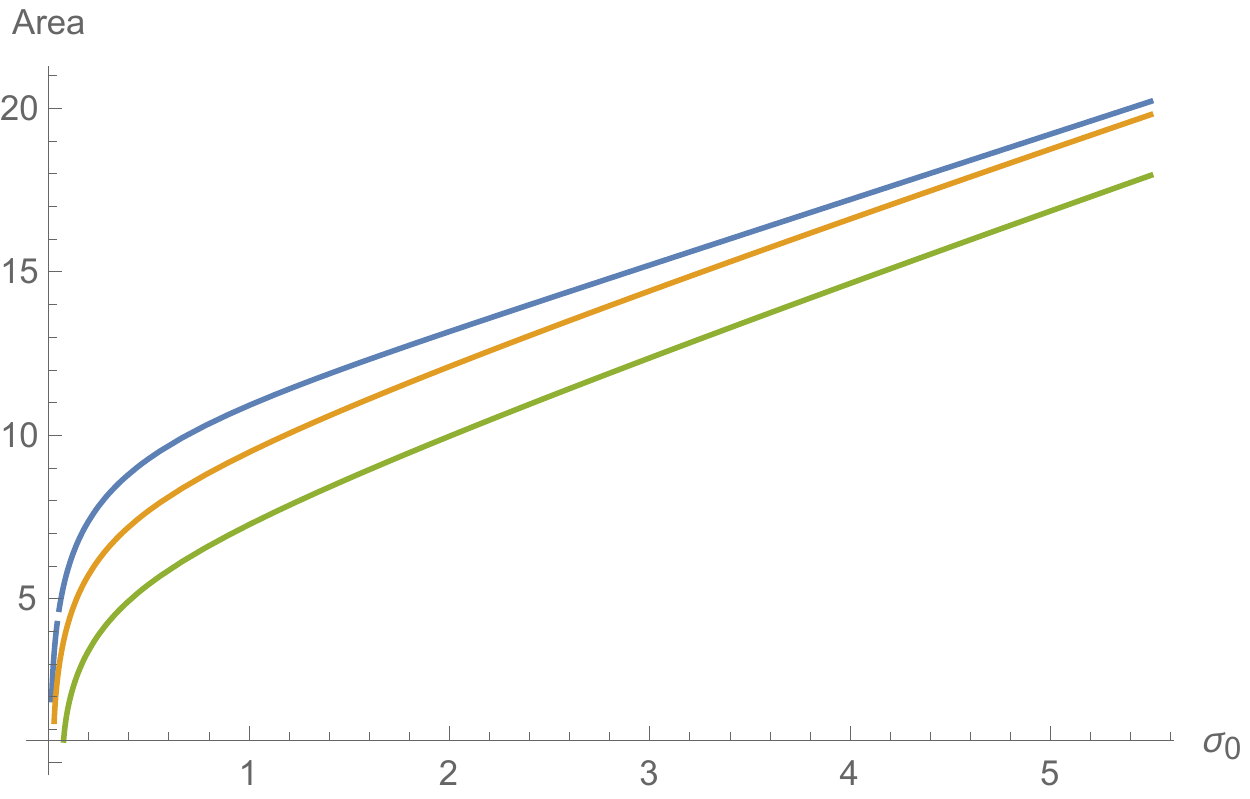}
\caption{Plot of the area of the HRT surface versus $R$, for three fixed values of $\frac{\tau_{0}}{R}$ (namely, $\frac{\tau_{0}}{R} = 0.1$ for the blue curve, $\frac{\tau_{0}}{R} = 0.9$ for the orange curve, and $\frac{\tau_{0}}{R} = 0.99$ for the green curve). The $r_{cutoff}$ has been set to $100$.}
\label{fig:BTZplot}
\end{figure}

In the remainder of this subsection, we set $L=1$ for convenience (but we will restore $L$ in the next section). We map to AdS-Rindler (otherwise known as the planar BTZ black hole), by the coordinate transformation
\begin{equation}
    x^{\pm} = \pm \sqrt{1 - r^{-2}} e^{\pm \sigma^{\pm}},
\end{equation}
\begin{equation}
    z = \frac{e^{\sigma}}{r},
\end{equation}
where $x^{\pm} \equiv t \pm x$ and $\sigma^{\pm} \equiv \tau \pm \sigma$. The AdS-Rindler metric is
\begin{equation}
    ds^{2} = -(r^{2} - 1)d\tau^{2} + \frac{dr^{2}}{r^{2} - 1} + r^{2}d\sigma^{2},
\end{equation}
and the image of the geodesic (\ref{zofu})--(\ref{xofu}) is found to be
\begin{eqnarray}
    \tau(\sigma) &=& \mathrm{arctanh} \bigg[ \nonumber \\
    && \frac{\sinh{\tau_R} \sinh{(\sigma-\sigma_L)} + \sinh{\tau_L} \sinh{(\sigma_{R} - \sigma)}} {\cosh{\tau_R} \sinh{(\sigma-\sigma_{L})} + \cosh{\tau_L} \sinh{(\sigma_{R} - \sigma)}} \bigg], \nonumber \\
    &&
\end{eqnarray}
\begin{eqnarray}
    r(\sigma) &=& \sinh{(\sigma_{R} - \sigma_{L})} \sqrt{\frac{\mathrm{csch}(\sigma-\sigma_{L}) \mathrm{csch}(\sigma_{R}-\sigma)}{2[\cosh{(\sigma_{R}-\sigma_{L})} - \cosh{(\tau_{R}-\tau_{L})}}}. \nonumber \\
    &&
\end{eqnarray}
In the expression above, $(\tau_{L},\sigma_{L})$ and $(\tau_{R},\sigma_{R})$ are the boundary coordinates of the two end points of the geodesic, respectively. Beyond this point, we will use translation symmetry to set $-\tau_{L} = \tau_{R} \equiv \tau_{0}$ and $-\sigma_{L} = \sigma_{R} \equiv R$. Using the explicit formulas for the geodesics above, we can write down the area of the HRT surface (i.e., the length of the geodesic). After some straightforward but tedious maths, we obtain
\begin{eqnarray}
    \mathrm{Area} &=& 2\sinh{(2 R)} \int_{r_{tip}}^{r_{cutoff}} \frac{dr}{r} \nonumber \\
    && \bigg[ \bigg( \cosh{2 R} - \frac{\sinh^{2}{2 R}}{r^{2}(\cosh{2 R} - \cosh{2\tau_{0}})} \bigg)^{2} - 1 \bigg]^{-1/2}, \nonumber\\
    &&
\end{eqnarray}
with $r_{tip}$ the $r$ coordinate of the tip of the HRT surface, given in terms of $R$ and $\tau_{0}$ by
\begin{equation}
    r_{tip} = \frac{\sqrt{2}\cosh{R}}{\sqrt{\cosh{2 R} - \cosh{2\tau_{0}}}}
\end{equation}
and $r_{cutoff}$ some large near-boundary cutoff radial coordinate. The two expressions above specify the area as a function of $R$ and $\tau_{0}$ or, equivalently as a function of $R$ and the ratio $\frac{\tau_{0}}{R}$. To check the projected volume law, we can plot numerically the area as a function of $R$, at various fixed values of the ratio $\frac{\tau_{0}}{R}$.  We show the plots in Fig. \ref{fig:BTZplot}.\\ 
As can be seen Fig. \ref{fig:BTZplot}, the area scales linearly with $R$, with a slope that is independent of the ratio $\frac{\tau_{0}}{R}$. In this way, we have checked numerically the projected volume law.

\section{A more general argument} \label{Sec:General}

In this section, we consider more general subregions. We will consider the $d=3$ case to keep the notation simple and generalize to arbitrary dimensions in Appendix \ref{app:ArbDimen}.

\subsection{The $d=3$ case}
We still work with static coordinates in the bulk, but with polar coordinates instead of Cartesian ones,
\begin{equation}
    ds^{2} = \frac{L^2}{z^2} \left[ -f dt^{2} + \frac{dz^{2}}{f} + d\rho^{2} + \rho^{2}d\phi^{2} \right],
\end{equation}
and we consider a boundary subregion described by $\rho = R \tilde{\rho} (\phi)$ and $t = R \tilde{t} (\phi)$, where $R$ is a scaling parameter, which we will take to be large. The boundary subregion needs not lie on a static time slice or even a boosted time slice, but we will require that it is symmetric under $(t,x,y) \rightarrow (-t,-x,-y)$.

The HRT surface is described by $\rho(z,\phi)$ and $t(z,\phi)$. The area functional is
\begin{equation}\label{area}
    \mathcal{A} = \int_{\delta}^{z_m} dz \int_{0}^{2\pi} d\phi \mathcal{L} = \int_{\delta}^{z_m} dz \int_{0}^{2\pi} d\phi \frac{L^2}{z^2} \sqrt{Q}
\end{equation}
with
\begin{equation}\label{Q}
    Q = -f(t' \rho_{,\phi} - \rho' t_{,\phi})^{2} + \rho^{2} (-ft'^{2} + \frac{1}{f} + \rho'^{2}) - t_{,\phi}^{2} + \frac{\rho_{,\phi}^{2}}{f}.
\end{equation}
The trick is to study the variation of $A$ with respect to $R$. By a standard Hamilton-Jacobi argument, we have that this variation is
\begin{equation}\label{HJ}
    \frac{d\mathcal{A}}{dR} = \int d\phi \left( \Pi_{\rho}^{z} \frac{d\rho}{dR} \bigg|^{z_m}_{\delta} + \Pi_{t}^{z} \frac{dt}{dR} \bigg|_{\delta}^{z_m} - \mathcal{H}(z_{m}) \frac{dz_{m}}{dR} \right),
\end{equation}
where $\Pi_{\rho}^{z} \equiv \frac{\partial \mathcal{L}}{\partial \rho'}$ and $\Pi_{t}^{z} \equiv \frac{\partial \mathcal{L}}{\partial t'}$ are the conjugate momenta. The prime denotes differentiation with respect to $z$. Explicitly,
\begin{align}
    \Pi^{z}_{\rho} &= \frac{L^2}{z^{2} \sqrt{Q}} \left[ f(t'\rho_{,\phi} - \rho' t_{,\phi}) t_{,\phi} + \rho^{2}\rho' \right],\\
    \Pi^{z}_{t} &= -\frac{L^2}{z^{2} \sqrt{Q}} \bigg[ f(t' \rho_{,\phi} - \rho' t_{,\phi}) \rho_{,\phi} + \rho^{2} f t'  \bigg],
\end{align}
and $\mathcal{H} \equiv \Pi^{z}_{\rho} \rho' + \Pi^{z}_{t} t' - \mathcal{L}$ is the Hamiltonian density. Explicitly,
\begin{equation}
    \mathcal{H} = \frac{L^2}{z^{2}\sqrt{Q}} \left( -\frac{\rho^2}{f} - \frac{\rho_{,\phi}^{2}}{f} + t_{,\phi}^{2} \right).
\end{equation}
In Eq. (\ref{HJ}), the first two terms on the right-hand side come from the change in the shape of the HRT surface as $R$ is varied, and the last term comes from the change in $z_{m}$ as $R$ is varied. Technically, the right-hand side of (\ref{HJ}) includes also the term $\int dz \int d\phi \partial_{\phi} \bigg( \Pi^{\phi}_{\rho} \frac{d\rho}{dR} + \Pi^{\phi}_{t} \frac{dt}{dR} \bigg)$, but this latter term vanishes by periodicity in $\phi$.

From the explicit expressions (\ref{Pizrho}) and (\ref{Pizt}), we see that those conjugate momenta vanish when evaluated at $z=z_{m}$. This is because partial derivatives of $\rho$ or $t$ with respect to $\phi$ vanish at the tip of the HRT surface, by regularity, and also because $\rho(z=z_{m})$ vanishes. For the same reasons, we see that the Hamiltonian density $\mathcal{H}$ vanishes when evaluated at $z=z_{m}$. So Eq. (\ref{HJ}) simplifies to
\begin{equation}\label{HJsimple}
    \frac{d\mathcal{A}}{dR} = -\int d\phi \left( \Pi_{\rho}^{z} \frac{d\rho}{dR} \bigg|_{\delta} + \Pi_{t}^{z} \frac{dt}{dR} \bigg|_{\delta} \right).
\end{equation}

\subsubsection{The near-boundary expansions}
Because the right-hand side above is evaluated at the near-boundary cutoff $z=\delta$, we see that it is enough to know how the HRT surface looks like near the boundary to know $\frac{dA}{dR}$. Let us expand $\rho(z,\phi)$ and $t(z,\phi)$ near the boundary (i.e., as series in $z$),
\begin{align}\label{nearboundaryrho}
    \rho &= R \tilde{\rho}(\phi) + c_{1}(R,\phi) z + c_{2}{(R,\phi)} z^{2} + \dots,\\
	\label{nearboundaryt}
    t &= R \tilde{t}(\phi) + d_{1}(R,\phi) z + d_{2}{(R,\phi)} z^{2} + \dots,
\end{align}
where we will refer to the coefficients $c_{i}$ and $d_{i}$ as the Fefferman-Graham (FG) coefficients. We claim that the first coefficients $c_{1}$ and $d_{1}$ vanish. To see this, we will need to solve the equations of motion perturbatively near the boundary. From the area functional (\ref{area}) in the $d=3$ case, we find the equations of motion,
\begin{eqnarray}
    && -\frac{2}{z}Q \frac{\partial Q}{\partial \rho'} - \frac{1}{2} \frac{\partial Q}{\partial z} \frac{\partial Q}{\partial \rho'} + Q \frac{\partial}{\partial z} \frac{\partial Q}{\partial \rho'} \nonumber \\
    &-& \frac{1}{2} \frac{\partial Q}{\partial \phi} \frac{\partial Q}{\partial \rho_{,\phi}} + Q \frac{\partial}{\partial \phi} \frac{\partial Q}{\partial \rho_{,\phi}} = 0,
\end{eqnarray}
\begin{eqnarray}
    && -\frac{2}{z}Q \frac{\partial Q}{\partial t'} - \frac{1}{2} \frac{\partial Q}{\partial z} \frac{\partial Q}{\partial t'} + Q \frac{\partial}{\partial z} \frac{\partial Q}{\partial t'} \nonumber \\
    &-& \frac{1}{2} \frac{\partial Q}{\partial \phi} \frac{\partial Q}{\partial t_{,\phi}} + Q \frac{\partial}{\partial \phi} \frac{\partial Q}{\partial t_{,\phi}} = 0.
\end{eqnarray}
If we plug the near-boundary expansions (\ref{nearboundaryrho}) and (\ref{nearboundaryt}) into the equations of motion above and demand that the equations of motion are satisfied at each order in $z$, we find that the order $z^{-1}$ yields
\begin{equation}\label{negativefirstorderEOM1}
    -\frac{2}{z} Q^{(0)} \left(\frac{\partial Q}{\partial \rho'} \right)^{(0)} = 0,
\end{equation}
\begin{equation}\label{negativefirstorderEOM2}
    -\frac{2}{z} Q^{(0)} \left(\frac{\partial Q}{\partial t'} \right)^{(0)} = 0,
\end{equation}
where the superscript $(0)$ means the term is zeroth order in $z$. We have
\begin{equation}
    Q^{(0)} = -(d_{1} R \tilde{\rho}_{,\phi} - c_{1} R \tilde{t}_{,\phi})^{2} + R^{2} \tilde{\rho}^{2} (1+c_{1}^{2}-d_{1}^{2}) - R^{2} \tilde{t}_{,\phi}^{2} + R^{2} \tilde{\rho}_{,\phi}^{2},
\end{equation}
\begin{equation}
    \left( \frac{\partial Q}{\partial \rho'} \right)^{(0)} = 2R^{2} [c_{1}(\tilde{\rho}^{2} - \tilde{t}_{,\phi}^{2}) + d_{1} \tilde{\rho}_{,\phi} \tilde{t}_{,\phi}],
\end{equation}
\begin{equation}
    \left( \frac{\partial Q}{\partial t'} \right)^{(0)} = 2R^{2} [c_{1} \tilde{\rho}_{,\phi} \tilde{t}_{,\phi} - d_{1} (\tilde{\rho}^{2} + \tilde{\rho}_{,\phi}^{2})].
\end{equation}
The solution to (\ref{negativefirstorderEOM1}) and (\ref{negativefirstorderEOM2}) is, as claimed,
\begin{equation}
    c_{1} = d_{1} = 0.
\end{equation}
We now plug the near-boundary expansions for $\rho$ and $t$ above into $\Pi^{z}_{\rho}$ and $\Pi^{z}_{t}$ and use the fact that the first-order coefficients vanish. We then find, in the $\delta \rightarrow 0$ limit
\begin{equation}\label{Pizrho}
    \Pi^{z}_{\rho} = \frac{3L^{2}R (d_{3} \tilde{\rho}_{,\phi} \tilde{t}_{,\phi} + c_{3} \tilde{\rho}^{2} - c_{3} \tilde{t}_{,\phi}^{2}) }{\sqrt{\tilde{\rho}^{2} - \tilde{t}_{,\phi}^{2} + \tilde{\rho}_{,\phi}^{2}}} + \dots,
\end{equation}
\begin{equation}\label{Pizt}
    \Pi^{z}_{t} = \frac{3L^{2}R (c_{3} \tilde{\rho}_{,\phi} \tilde{t}_{,\phi} - d_{3} \tilde{\rho}^{2} - d_{3} \tilde{\rho}_{,\phi}^{2}) }{\sqrt{\tilde{\rho}^{2} - \tilde{t}_{,\phi}^{2} + \tilde{\rho}_{,\phi}^{2}}} + \dots,
\end{equation}
where the $\dots$ stands for terms that either diverge or go to zero as $\delta \rightarrow 0$. In other words, we have only written down the term that remains finite in the $\delta \rightarrow 0$ limit, because intuitively we expect this finite contribution to give rise to volume-law entanglement entropy.  To further stress the point that we are interested only in the finite contributions to the entanglement entropy, we are not including divergent terms associated with the vacuum entropy that should be regularized.  We note also that the quantity $\tilde{\rho}^{2} - \tilde{t}_{,\phi}^{2} + \tilde{\rho}_{,\phi}^{2}$ is positive since the subregion is spacelike.

\subsubsection{The large-$R$ expansion}
In the large-$R$ limit, we guess that $\rho$ and $t$ can be expanded in $1/R$ as
\begin{align}\label{largeRrho}
    \rho &= R \tilde{\rho}{(\phi)} - \rho_{0}{(z,\phi)} + \frac{\rho_{1}(z,\phi)}{R} + \frac{\rho_{2}(z,\phi)}{R^{2}} + \dots,\\
	\label{largeRt}
    t &= R\tilde{t}{(\phi)} - t_{0}{(z,\phi)} + \frac{t_{1}(z,\phi)}{R} + \frac{t_{2}(z,\phi)}{R^{2}} + \dots
\end{align}
(without noninteger powers of $R$). Then $Q$ can be expanded in a series in $1/R$ without any noninteger powers,
\begin{equation}
    Q = R^{2} Q^{(-2)} + RQ^{(-1)} + Q^{(0)} + \frac{1}{R} Q^{(1)} + \frac{1}{R^2} Q^{(2)} + \dots
\end{equation}
and similarly for the area functional
\begin{equation}
    \mathcal{A} = R \mathcal{A}^{(-1)} + \mathcal{A}^{(0)} + \frac{1}{R} \mathcal{A}^{(1)} + \frac{1}{R^2} \mathcal{A}^{(2)} + \dots~.
\end{equation}
We treat the expansions above as an ansatz. For the purpose of the volume law, we will need only the coefficients $\rho_{0}$ and $t_{0}$. In Appendix \ref{app:largeR}, we take a closer look at the general structure of the higher-order perturbation theory in $1/R$, as well as check explicitly that the ansatz above matches with the exact HRT surface in the strip case.  The area functional to leading order in $1/R$ takes the form
\begin{equation}
    R \mathcal{A}^{(-1)} = R \int_{\delta}^{z_m} dz \int_{0}^{2\pi} d\phi \frac{L^2}{z^2} \sqrt{Q^{(-2)}},
\end{equation}
with
\begin{eqnarray}
    Q^{(-2)} &=& -f(-t_{0}' \tilde{\rho}_{,\phi} + \rho_{0}' \tilde{t}_{,\phi})^{2} + \tilde{\rho}^{2} \left[ -f(t_{0}')^{2} + \frac{1}{f} + (\rho_{0}')^{2} \right] \nonumber \\
    &-& \tilde{t}_{,\phi}^{2} + \frac{\tilde{\rho}_{,\phi}^{2}}{f}.
\end{eqnarray}
The equations of motion take the form
\begin{equation}\label{aofphi}
    \frac{\partial \mathcal{L}}{\partial t_{0}'} = \frac{L^2}{z^2} \frac{f}{\sqrt{Q^{(-2)}}} \left[ \rho_{0}' \tilde{\rho}_{,\phi} \tilde{t}_{,\phi} -t_{0}' (\tilde{\rho}_{,\phi}^{2} + \tilde{\rho}^{2}) \right] = a(\phi),
\end{equation}
\begin{equation}\label{bofphi}
    \frac{\partial \mathcal{L}}{\partial \rho_{0}'} = \frac{L^2}{z^2} \frac{1}{\sqrt{Q^{(-2)}}} \left[ f \tilde{\rho}_{,\phi} \tilde{t}_{,\phi} t_{0}' + \rho_{0}' (\tilde{\rho}^{2} - f \tilde{t}_{,\phi}^{2}) \right] = b(\phi),
\end{equation}
where $a(\phi)$ and $b(\phi)$ are some functions of $\phi$ only, and not of $z$. Taking the ratio of the two equations above, and solving for $t_{0}'$, we find
\begin{equation}
    t_{0}' = F(\phi) \rho_{0}',
\end{equation}
with
\begin{equation}
    F(\phi) = \frac{\tilde{\rho}_{,\phi}\tilde{t}_{,\phi} b(\phi) - \left( \frac{\tilde{\rho}^{2}}{f} - \tilde{t}_{,\phi}^{2} \right) a(\phi)}{a(\phi)\tilde{\rho}_{,\phi}\tilde{t}_{,\phi} + b(\phi) (\tilde{\rho}_{,\phi}^{2} + \tilde{\rho}^{2}) }.
\end{equation}
Evaluating the conjugate momenta at the tip of the HRT surface, and using the equation above together with regularity at the tip $(\rho_{0}' \rightarrow \infty)$ and the fact that $f(z_{m})$ is close to zero for large $R$, we then conclude that
\begin{align}
    a(\phi) &= 0,\\
    b(\phi) &= \frac{L^2}{z_{m}^{2}} \tilde{\rho},
\end{align}
and we find that $\rho_{0}$ satisfies the differential equation
\begin{equation}
 \frac{A(\phi,z)\rho_{0}'}{\sqrt{B(\phi,z) \rho_{0}'^{2} + C(\phi,z)}} = \frac{L^2}{z_{m}^{2}} \tilde{\rho},
\end{equation}
with
\begin{align}
    A(\phi,z) &= \frac{L^2}{z^2} \tilde{\rho}^{2} (1 - \xi),\\
    B(\phi,z) &= \tilde{\rho}^{2} (1-\xi),\\
    C(\phi,z) &= \frac{\tilde{\rho}^{2} + \tilde{\rho}_{,\phi}^{2}}{f} (1-\xi),
\end{align}
where we defined
\begin{equation}
    \xi \equiv \frac{f \tilde{t}_{,\phi}^{2}}{\tilde{\rho}^{2} + \tilde{\rho}_{,\phi}^{2}}.
\end{equation}
Solving for $\rho_{0}'$, we find
\begin{equation}\label{90}
    \rho_{0}' = \sqrt{1 + \frac{\tilde{\rho}_{,\phi}^{2}}{\tilde{\rho}^{2}}} \frac{1}{\sqrt{f}} \frac{z^{2}}{\sqrt{z_{m}^{4} (1-\xi) - z^{4}}}.
\end{equation}

At this point, a few remarks are in order about the nature of our large-$R$ expansion. We note that, in principle, $z_{m}$ in the expression above is itself a function of $R$ (the size of the boundary subregion determines how deep the surface penetrates into the bulk), so the $\rho_{0}$ above is not exactly a quantity zeroth order in $1/R$. It is tempting to resolve this subtlety by replacing $z_{m}$ by its zeroth-order part, which is the horizon location $z_{h} = 1$. Doing so, however, is unsatisfactory because the size of the subregion is infinite when $z_{m}$ is at the horizon (as can be seen by integrating $\rho_{0}'$ and checking that there is a divergence near the horizon). It would be preferable to deal with large but finite subregions.

We prefer to think about the large-$R$ expansion in this way: we hypothesize that there is a way to split the $R$ dependence of $\rho(z,\phi)$ and $t(z,\phi)$ into a dependence through $z_{m}$ and a dependence not through $z_{m}$. Then the large-$R$ expansion consists of expanding the $R$ dependence not through $z_{m}$, while leaving $z_{m}$ unexpanded. In Appendix \ref{app:largeR}, we illustrate more concretely how such $z_{m}$ arises in the case of strips.

For the purpose of the volume law, however, it seems harmless to replace $z_{m}$ by the horizon value $1$, so we set $z_{m}$ to $1$ now.

Expanding in $z$ around $z=0$, we find the leading-order term to be
\begin{equation}
    \rho_{0}' = \frac{1}{\tilde{\rho}} \frac{\tilde{\rho}^{2} + \tilde{\rho}_{,\phi}^{2}}{\sqrt{\tilde{\rho}^{2} + \tilde{\rho}_{,\phi}^{2} - \tilde{t}_{,\phi}^{2}}} z^{2} + \dots ~.
\end{equation}
Integrating, we find the coefficient $c_{3}$ in the large-$R$ limit to be
\begin{equation}\label{c3}
    c_{3} = -\frac{1}{3 \tilde{\rho}} \frac{\tilde{\rho}^{2} + \tilde{\rho}_{,\phi}^{2}}{\sqrt{\tilde{\rho}^{2} + \tilde{\rho}_{,\phi}^{2} - \tilde{t}_{,\phi}^{2}}},
\end{equation}
and the coefficient $d_{3}$ is found to be
\begin{equation}\label{d3}
    d_{3} = -\frac{1}{3 \tilde{\rho}} \frac{\tilde{\rho}_{,\phi} \tilde{t}_{,\phi}}{\sqrt{\tilde{\rho}^{2} + \tilde{\rho}_{,\phi}^{2} - \tilde{t}_{,\phi}^{2}}}.
\end{equation}

\subsubsection{The volume law}
Plugging back the answers for $c_{3}$ and $d_{3}$ in (\ref{c3}) and (\ref{d3}) into the conjugate momenta $\Pi^{z}_{\rho}$ and $\Pi^{z}_{t}$ in (\ref{Pizrho}), (\ref{Pizt}), we obtain:
\begin{align}
    \Pi^{z}_{\rho} &= -L^{2} R \tilde{\rho},\\
    \Pi^{z}_{t} &= 0.
\end{align}
Plugging the two results above back into $\frac{dA}{dR}$, we then find
\begin{equation}\label{dAdRfinal}
    \frac{d\mathcal{A}}{dR} = \int_{0}^{2\pi} d\phi L^{2} R \tilde{\rho}(\phi)^{2}.
\end{equation}
Finally, consider the volume of the \textit{projection} of the subregion onto a static time slice. The boundary of that projected subregion is described by $\rho(\phi) = R \tilde{\rho}(\phi)$ (with $t=0$). The volume of that projected subregion is found to be:
\begin{equation}
    \mathrm{Volume} = \int_{0}^{2\pi} d\phi \int_{0}^{R \tilde{\rho}(\phi)} \rho d\rho = \frac{1}{2} R^{2} \int_{0}^{2\pi} d\phi \tilde{\rho}{(\phi)}^{2}.
\end{equation}
Differentiating with respect to $R$,
\begin{equation}
    \frac{d}{dR} \mathrm{Volume} = R \int_{0}^{2\pi} d\phi \tilde{\rho}{(\phi)}^{2}.
\end{equation}
Comparing the above with (\ref{dAdRfinal}) for $\frac{d\mathcal{A}}{dR}$, we then see that $\frac{dA}{dR}$ is proportional to $\frac{d \mathrm{Volume}}{R}$, up to a factor of $L^{2}$, which is an entropy density. In other words, the area $\mathcal{A}$ of the HRT surface indeed scales with the volume of the projected subregion.

\section{Conclusion}\label{Sec:Conclusion}
In this paper, we have taken the first steps toward establishing a version of volume-law entanglement entropy, by using HRT surfaces in the eternal black brane spacetime which are anchored at boundary subregions which break staticity of the thermal states. We consider strips as well as general subregions which are reflection symmetric. We emphasize that this work is not a rigorous proof, but the findings agree with intuition from field theory as well as more rigorous computations in field theory.

A special case of our result of particular physical interest is that of the boosted black brane, which is the gravity dual to nonequilibrium steady states (or NESSs). In particular, the entanglement structure of NESSs was studied in \cite{Mahajan:2016tem} (without using holography) or in \cite{Bhaseen:2013ypa} (using the boosted black brane).

A natural direction for future work is to generalize to subregions without reflection symmetry. Then the techniques used here do not apply since reflection symmetry guarantees that the tip of the HRT surface stays at the same boundary coordinates as $R$ is varied.  This in turn allows for the Hamilton-Jacobi argument at the beginning of Sec. \ref{Sec:General} that relates $\frac{dA}{dR}$ to the near-boundary behavior of the HRT surface. In the absence of reflection symmetry, we need a more general technique, perhaps a ``matching technique'' similar to the one used in \cite{Liu:2012eea, Liu:2013una}.  A different approach would be to use Gauss' identity for the first variation of the area as in Appendix A of \cite{Engelhardt:2019hmr}.  The primary difficulty in Gauss's identity approach is computing the vector field $\eta^a$ of \cite{Engelhardt:2019hmr} explicitly in the large-$R$ limit which seems nontrivial at this time.

A different direction for future work is a generalization to other bulk spacetimes. In particular, it would be interesting to study the volume law in the context of the hydrodynamic black hole solution \cite{Bhattacharyya:2007vjd, Hubeny:2010wp}. Here again, a more general technique than the one employed in this paper will be necessary.  We also expect that there will be corrections to the projected volume law in this case due to the fact that entropy can be generated in these scenarios.  This can be seen from the derivative expansion of a hydrodynamic black hole where the zeroth-order term is a boosted black hole. The projected volume law should be true to zeroth order in this case with subsequent corrections.

\begin{acknowledgments}
The work of T.G. is supported by the National Science Foundation under Grant No. PHY-1914679. P.N. acknowledges support from Israel Science Foundation Grant No. 447/17 for the work done in Sec. \ref{Sec:FieldTheory} and from U.S. National Science Foundation Grant No. PHY-1820734 at CUNY for the work done in Sec. \ref{Sec:Strips}.

The authors are very thankful for Brian Swingle, Mark
Mezei, and Matthew Headrick for useful discussions, and
especially for help with the Sec. \ref{Sec:FieldTheory}. We thank
Josiah Couch for collaboration during the early stage of
the project.
\end{acknowledgments}

\appendix

\section{More on the large-$R$ expansion}\label{app:largeR}

Next, we perform a consistency check of the large-$R$ expansion by showing that it is consistent with the exact HRT surface for a strip in $3+1$ bulk dimensions and also quantify the error of the large-$R$ approximation in this case. The exact shape of the HRT surface in that case is given by Eqs. (\ref{xstrip}) and (\ref{tstrip}), reproduced here for convenience,
\begin{align}\label{A4}
    x(z) &= R + \int_{0}^{z} dz' \frac{p_x}{\sqrt{p_{t}^{2} + f(z') \left( \frac{L^4}{z'^4} - p_{x}^{2} \right)}},\\
	\label{A5}
    t(z) &= t_{0} - \int_{0}^{z} dz' \frac{p_t}{f(z')} \frac{1}{\sqrt{p_{t}^{2} + f(z') \left( \frac{L^4}{z'^4} - p_{x}^{2} \right)}}.
\end{align}
To compare with the large-$R$ approximation, we first need to convert from Cartesian coordinates to polar coordinates using the well-known relation $x = \rho \cos{\phi}$ and $y = \rho \sin{\phi}$. Equation (\ref{A4}) becomes
\begin{equation}
    \rho(z,\phi) = R \tilde{\rho}(\phi) + \int_{0}^{z} dz' \frac{p_x \tilde{\rho}(\phi)}{\sqrt{p_{t}^{2} + f(z') \left( \frac{L^4}{z'^4} - p_{x}^{2} \right)}},
\end{equation}
with $\tilde{\rho}(\phi) = \sec{\phi}$. Let us take the derivative with respect to $z'$,
\begin{equation}
    \rho'(z, \phi) = \frac{p_{x} \tilde{\rho}(\phi)}{\sqrt{p_{t}^{2} + f(z) \left( \frac{L^4}{z^4} - p_{x}^{2} \right)}}.
\end{equation}
The expression above is exact. In the large-$R$ limit, as is argued in Sec. \ref{Sec:Strips}, we have $p_{t} \approx 0$ and $p_{x} \approx -\frac{L^2}{z_{m}^{2}}$, and the above becomes
\begin{equation}
    \rho'(z, \phi) \approx -\tilde{\rho}{(\phi)} \frac{1}{\sqrt{f}} \frac{z^2}{\sqrt{z_{m}^{4} - z^{4}}}.
\end{equation}
Note that we do not expand $z_m$ into a series in $1/R$, which is in accordance with the rule described in Sec. \ref{Sec:General}.  The point now is to check that the above is the same as $-\rho_{0}'$, with $\rho_{0}'$ given by Eq. (\ref{90}), when we specialize that equation to a strip. We note that the function $\tilde{t}(\phi)$ is independent of $\phi$ for a strip, so the quantity $\xi$ in (\ref{90}) vanishes. Furthermore, the factor $\sqrt{1 + \frac{\tilde{\rho}_{,\phi}^{2}}{\tilde{\rho}^{2}}}$ in that equation simplifies to $\tilde{\rho}$ when we take $\tilde{\rho}(\phi) = \sec{\phi}$ as is appropriate for a strip. Then we see that the equation above is indeed the same as $-\rho_{0}'$.

We can similarly check that $t'$ as obtained from (\ref{A5}) matches with $-t_{0}'$ as obtained from (\ref{81}). Both expressions are easily seen to vanish.

Generically, these calculations show that we are not missing any ``intermediate'' terms in the large-$R$ expansion such as $\sqrt{R}$, $R^{1/4}$, etc., at least for strips. Perhaps the absence of noninteger powers of $1/R$ can be argued more carefully from results such as \cite{Grover:2011fa}.

Next, we take a closer look at the general structure of the higher-order perturbation theory in $1/R$. Consider first the term $\mathcal{A}^{(0)}$ in the action. We find that it is given in terms of $Q$ by
\begin{equation}
    \mathcal{A}^{(0)} = \frac{1}{2} \int dz \int d\phi \frac{L^{2}}{z^{2}} \frac{Q^{(-1)}}{\sqrt{Q^{(-2)}}},
\end{equation}
with
\begin{equation}
    Q^{(-1)} = F_{1}{(z,\phi)} \rho_{1}' + F_{2}{(z,\phi)} t_{1}' + F_{3}{(z,\phi)}
\end{equation}
and 
\begin{equation}
    F_{1}{(z,\phi)} = -2\tilde{\rho}^{2}\rho_{0}' + 2f\tilde{t}_{,\phi}(\tilde{t}_{,\phi}\rho_{0}' - \tilde{\rho}_{,\phi}t_{0}'),
\end{equation}
\begin{equation}
    F_{2}{(z,\phi)} = 2f \tilde{\rho}^{2} t_{0}' - 2f \tilde{\rho}_{,\phi} (\tilde{t}_{,\phi}\rho_{0}' - \tilde{\rho}_{,\phi} t_{0}'),
\end{equation}
\begin{eqnarray}
    F_{3}{(z,\phi)} &=& 2t_{0,\phi}\tilde{t}_{,\phi} -2\rho_{0}\tilde{\rho} \left( \frac{1}{f} - ft_{0}'^{2} + \rho_{0}'^{2} \right) - \frac{2}{f} \rho_{0,\phi} \tilde{\rho}_{,\phi} \nonumber \\
    &+& 2f (\tilde{t}_{,\phi}\rho_{0}' - \tilde{\rho}_{,\phi} t_{0}') (-t_{0}'\rho_{0,\phi} + t_{0,\phi} \rho_{0}').
\end{eqnarray}
Note that, in the action $\mathcal{A}^{(0)}$, the $F_{1}$, $F_{2}$, $F_{3}$, and $Q^{(-2)}$ are known functions. Because the action above is linear in the first derivatives $\rho_{1}'$ and $t_{1}'$, we might worry that there is an inconsistency, unless the quantities
\begin{equation}
    \frac{L^2}{z^2} \frac{F_{1}(z,\phi)}{\sqrt{Q^{(-2)}}}
\end{equation}
and
\begin{equation}
    \frac{L^2}{z^2} \frac{F_{2}(z,\phi)}{\sqrt{Q^{(-2)}}}
\end{equation}
turn out to be independent of $z$ (which is what the equations of motion predict). And indeed they are, because they are equal to $-2 b(\phi)$ and $-2 a(\phi)$ with $a(\phi)$ and $b(\phi)$ as defined in Eqs. (\ref{aofphi}) and (\ref{bofphi}). So the perturbation theory is self-consistent, but we have learned that we need to go to the next order in the action to find $\rho_{1}$ and $t_{1}$.

So then, consider the term $\mathcal{A}^{(1)}$ in the action. It is given in terms of $Q$ by
\begin{equation}
    \mathcal{A}^{(1)} = \int dz \int d\phi \frac{L^2}{z^2} \frac{1}{Q^{(-2)}{}^{3/2}} \left( \frac{1}{2} Q^{(0)}Q^{(-2)} - \frac{1}{8} Q^{(-1)}{}^{2} \right),
\end{equation}
where $Q^{(-2)}$ and $Q^{(-1)}$ have previously been written down, and $Q^{(0)}$ is found to be
\begin{eqnarray}
Q^{(0)} &=& -t_{0,\phi}^{2} - 2t_{1,\phi}\tilde{t}_{,\phi} - 4\rho_{0}\tilde{\rho} (ft_{0}' t_{1}' - \rho_{0}'\rho_{1}') \nonumber \\
&+& \tilde{\rho}^{2} (-f t_{1}'^{2} + 2ft_{0}'t_{2}' + \rho_{1}'^{2} - 2\rho_{0}'\rho_{2}') \nonumber \\
&+& \left( \frac{1}{f} - f t_{0}'^{2} + \rho_{0}'^{2} \right) (\rho_{0}^{2} + 2\rho_{1} \tilde{\rho}) \nonumber \\
&-& f(t_{0}' \rho_{0,\phi} - t_{0,\phi}\rho_{0}' - \tilde{t}_{,\phi}\rho_{1}' + t_{1}'\tilde{\rho}_{,\phi})^{2} \nonumber \\
&+& 2f (\tilde{t}_{,\phi}\rho_{0}' - t_{0}' \tilde{\rho}_{,\phi}) ( t_{1}'\rho_{0,\phi} - t_{1,\phi}\rho_{0}' \nonumber \\
&+& t_{0}'\rho_{1,\phi} - t_{0,\phi}\rho_{1}' + \tilde{t}_{,\phi}\rho_{2}' - t_{2}' \tilde{\rho}_{,\phi} ) \nonumber \\
&+& \frac{\rho_{0,\phi}^{2} + 2\rho_{1,\phi} \tilde{\rho}_{,\phi}}{f}.
\end{eqnarray}

By inspection, the equations of motion for $\rho_{1}$ and $t_{1}$ do not involve $\rho_{2}$ and $t_{2}$, so we can solve for $\rho_{1}$ and $t_{1}$ from those equations of motion. This is another check that the perturbation theory is consistent. The differential equations that determine $\rho_{1}$ and $t_{1}$ have the form of a coupled system of second-order ordinary differential equations, the solution of which can be written down analytically in integral form but we will not do so because it is not illuminating. Also, the equations of motion from the action $\mathcal{A}^{1}$ for $\rho_{2}$ and $t_{2}$ turn out to be the same consistency check as the one we have seen coming from $\mathcal{A}^{(0)}$. This is another piece of evidence that the perturbation is consistent.

\section{Volume-law scaling in arbitrary dimensions} \label{app:ArbDimen}

In this appendix, we generalize the computation in Sec. \ref{Sec:General} to arbitrary $d$. The metric is now
\begin{equation}
	ds^{2} = \frac{L^2}{z^2} \left[ -f(z) dt^{2} + \frac{dz^{2}}{f(z)} + d\rho^{2} + \rho^{2}d\Omega_{d-2}^{2} \right],
\end{equation}
with
\begin{equation}
	f(z) = 1 - z^{d},
\end{equation}
and we consider a boundary subregion described by $\rho = R \tilde{\rho} (\phi)$ and $t = R \tilde{t} (\phi)$, where $R$ is a scaling parameter that we will take to be large. The boundary subregion needs not lie on a static time slice or even a boosted time slice, but -- as in the case of  strips in Sec. \ref{Sec:General} -- we will require that it is symmetric under the reflection $(t,x,y) \rightarrow (-t,-x,-y)$. We write the metric on the $(d-2)$-sphere as
\begin{equation}
	d\Omega_{d-2}^{2} = \sum_{i=1}^{d-2} g_{i} d\theta_{i}^{2},
\end{equation}
with $g_{1} = 1$, $g_{2} = \sin^{2}{\theta_{1}}$, $g_{3} = \sin^{2}{\theta_{1}} \sin^{2}{\theta_{2}}$, etc.

The HRT surface is described by two functions $\rho(z,\theta_{i})$ and $t(z,\theta_{i})$. The induced metric on the surface is
\begin{eqnarray}
	ds^{2} &=& \frac{L^2}{z^2} \bigg[ A dz^{2} + 2B_{i} dz d\theta^{i} + D_{ij} d\theta^{i} d\theta^{j} \bigg],
\end{eqnarray}
where 
\begin{align}
	A &= -f(z) t'^{2} + \frac{1}{f(z)} + \rho'^{2},\\
	B_{i} &= -ft' t_{,i} + \rho' \rho_{,i},\\
	D_{ij} &= -ft_{,i}t_{,j} + \rho_{,i}\rho_{,j} + \rho^{2} g_{i} \delta_{ij},
\end{align}
and $t_{,i} \equiv \frac{\partial t}{\partial \theta^{i}}$. The area functional is
\begin{equation}
	\mathcal{A} = \int_{\delta}^{z_m} \int d\Omega_{d-2} \left( \frac{L}{z} \right)^{d-1} \sqrt{Q},
\end{equation}
with
\begin{eqnarray}
	Q &=& \rho^{2(d-2)} \det(d\Omega_{d-2}^{2}) \bigg \{ -f t'^{2} + \frac{1}{f} + \rho'^{2} \nonumber \\
	&+& \frac{1}{\rho^2} \sum_{i} \frac{1}{g_i} \bigg[ \frac{\rho_{,i}^{2}}{f} - t_{,i}^{2} - f(\rho' t_{,i} - t' \rho_{,i})^{2} \bigg] \nonumber \\
	&-& \frac{1}{\rho^{4}} \sum_{i<j} \frac{(\rho_{,i}t_{,j}-t_{,i}\rho_{,j})^2}{g_{i}g_{j}} \bigg \}.
\end{eqnarray}
The $\frac{d\mathcal{A}}{dR}$ is now given by (again by using the Hamilton-Jacobi argument)
\begin{equation}\label{HJarbitraryd}
	\frac{d\mathcal{A}}{dR} = -\int d\Omega_{d-2} \left( \Pi^{z}_{\rho} \frac{d\rho}{dR} \bigg|_{\delta} + \Pi^{z}_{t} \frac{dt}{dR} \bigg|_{\delta} \right),
\end{equation}
where the conjugate momenta are
\begin{eqnarray}
	\Pi^{z}_{\rho} &=& \left( \frac{L}{z} \right)^{d-1} \frac{1}{\sqrt{Q}} \rho^{2(d-2)} \mathrm{det}{(d\Omega_{d-2}^{2})} \nonumber \\
	&& \bigg[ \rho' - \frac{1}{\rho^2} \sum_{i} \frac{f t_{,i}}{g_i} (\rho' t_{,i} - t' \rho_{,i})  \bigg],
\end{eqnarray}
\begin{eqnarray}
	\Pi^{z}_{t} &=& \left( \frac{L}{z} \right)^{d-1} \frac{1}{\sqrt{Q}} \rho^{2(d-2)} \mathrm{det}{(d\Omega_{d-2}^{2})} \nonumber \\
	&& \bigg[ -f t' + \frac{1}{\rho^2} \sum_{i} \frac{f \rho_{,i}}{g_i} (\rho' t_{,i} - t' \rho_{,i})  \bigg].
\end{eqnarray}
We note that the right-hand side of (\ref{HJarbitraryd}) also includes the term $\int dz \int d\Omega_{d-2} \partial_{i} \left( \Pi^{i}_{\rho} \frac{d\rho}{dR} + \Pi^{i}_{t} \frac{dt}{dR} \right)$. But we can easily see that this term vanishes by virtue of $\Pi^{i}_{\rho}$ vanishing when $\theta^{i}$ is evaluated at $0$ or $\pi$.

Next, we plug the near-boundary expansions (\ref{nearboundaryrho}) and (\ref{nearboundaryt}) into the above. Let us assume that, in the large-$R$ limit, the leading terms in the expansion for $\rho'$ and $t'$ are the $d$th FG coefficients,
\begin{equation}
	\lim_{R \rightarrow \infty} \rho' = c_{d}{(R,\phi)} d z^{d-1} + \dots,
\end{equation}
\begin{equation}
	\lim_{R \rightarrow \infty} t' = d_{d}{(R,\phi)} d z^{d-1} + \dots,
\end{equation}
and that the FG coefficients of lower order than the $d$th ones are $1/R$ suppressed. We will justify this assumption \emph{a posteriori} when we look at the large-$R$ equation of motion later. The conjugate momenta can then be written in terms of the FG coefficients as
\begin{eqnarray}
	\Pi^{z}_{\rho} &=& L^{d-1} R^{d-2} \tilde{\rho}^{d-2} \sqrt{\mathrm{det}{(d\Omega_{d-2}^{2})}} \frac{d}{\sqrt{\tilde{Q}}} \nonumber \\
	&\times& \bigg[ c_{d}\tilde{\rho}^{2} - \sum_{i} \frac{\tilde{t}_{,i}}{g_i} (c_{d} \tilde{t}_{,i} - d_{d}\tilde{\rho}_{,i}) \bigg] \nonumber \\
	&+& \dots,
\end{eqnarray}
\begin{eqnarray}
	\Pi^{z}_{t} &=& L^{d-1} R^{d-2} \tilde{\rho}^{d-2} \sqrt{\mathrm{det}{(d\Omega_{d-2}^{2})}} \frac{d}{\sqrt{\tilde{Q}}} \nonumber \\
	&\times& \bigg[ -d_{d}\tilde{\rho}^{2} + \sum_{i} \frac{\tilde{\rho}_{,i}}{g_i} (c_{d} \tilde{t}_{,i} - d_{d}\tilde{\rho}_{,i}) \bigg] \nonumber \\
	&+& \dots,
\end{eqnarray}
where
\begin{equation}
	\tilde{Q} = \tilde{\rho}^{4} + \tilde{\rho}^{2} \sum_{i} \frac{(\tilde{\rho}_{,i}^{2} - \tilde{t}_{,i}^{2})}{g_i} - \sum_{i<j} \frac{(\tilde{\rho}_{,i} \tilde{t}_{,j} - \tilde{t}_{,i} \tilde{\rho}_{,j})^2}{g_{i}g_{j}}.
\end{equation}
We note that $\tilde{Q}$ is a positive-definite quantity due to spacelikeness of the boundary subregion. [Indeed, if we compute the metric induced by $d$-dimensional Minkowski space onto the surface $\rho = \tilde{\rho}{(\theta_{i})}$, $t = \tilde{t}{(\theta_{i})}$, we find that the determinant of the induced metric is proportional to $\tilde{Q}$.]

Next, in order to obtain the coefficients $c_{d}$ and $d_{d}$ that appear above, we have to solve the equations of motion in the large-$R$ limit. We guess that the expansions of $\rho$ and $t$ in $1/R$ are
\begin{align}
	\rho &= R \tilde{\rho}{(\theta_{i})} - \rho_{0}{(z,\theta_{i})} + \mathcal{O}{(1/R)},\\
	t &= R \tilde{t}{(\theta_{i})} - t_{0}{(z,\theta_{i})} + \mathcal{O}{(1/R)}.
\end{align}
The area functional at large $R$ is
\begin{equation}
	\mathcal{A} = R^{d-2} \int_{\delta}^{z_m} dz \int d\Omega_{d-2} \left( \frac{L}{z} \right)^{d-1} \sqrt{Q''},
\end{equation}
with
\begin{eqnarray}
	Q'' &=& \tilde{\rho}^{2(d-2)} \mathrm{det}{(d\Omega_{d-2}^{2})} \bigg \{ -f t_{0}'^{2} + \frac{1}{f} + \rho_{0}'^{2} \nonumber \\
	&+& \frac{1}{\tilde{\rho}^{2}} \sum_{i} \frac{1}{g_i} \bigg[ \frac{\tilde{\rho}_{,i}^{2}}{f} - \tilde{t}_{,i}^{2} - f(t_{0}' \tilde{\rho}_{,i} - \rho_{0}' \tilde{t}_{,i})^{2} \bigg] \nonumber \\
	&-& \frac{1}{\tilde{\rho}^{4}} \sum_{i<j} \frac{(\tilde{\rho}_{,i}\tilde{t}_{,j} - \tilde{t}_{,i}\tilde{\rho}_{,j})^{2}}{g_{i}g_{j}} \bigg \}.
\end{eqnarray}
The equations of motion are
\begin{eqnarray}
	&& -\left( \frac{L}{z} \right)^{d-1} \frac{f}{\sqrt{Q''}} \tilde{\rho}^{2(d-2)} \mathrm{det}{(d\Omega_{d-2})} \nonumber \\
	&\times& \bigg[ t_{0}' + \frac{1}{\tilde{\rho}^{2}} \sum_{i} \frac{\tilde{\rho}_{,i}}{g_i} (t_{0}' \tilde{\rho}_{,i} - \rho_{0}' \tilde{t}_{,i}) \bigg] = a{(\theta^{i})},
\end{eqnarray}
\begin{eqnarray}
	&& \left( \frac{L}{z} \right)^{d-1} \frac{1}{\sqrt{Q''}} \tilde{\rho}^{2(d-2)} \mathrm{det}{(d\Omega_{d-2})} \nonumber \\
	&\times& \bigg[ \rho_{0}' + \frac{f}{\tilde{\rho}^{2}} \sum_{i} \frac{\tilde{t}_{,i}}{g_i} (t_{0}' \tilde{\rho}_{,i} - \rho_{0}' \tilde{t}_{,i}) \bigg] = b{(\theta^{i})},
\end{eqnarray}
for some functions $a$ and $b$ of the $(d-2)$ angles. Taking the ratio of the two equations above, and solving for $t_{0}'$, we find
\begin{equation}\label{81}
	t_{0}' = G(\theta^{i}) \rho_{0}',
\end{equation}
with
\begin{equation}
	G = \frac{b(\theta^{i}) \sum_{i} \frac{\tilde{t}_{,i}\tilde{\rho}_{,i}}{g_{i}} - a(\theta^{i}) \left( \frac{\tilde{\rho}^{2}}{f} - \sum_{i} \frac{\tilde{t}_{,i}^{2}}{g_i} \right)}{a(\theta^{i}) \sum_{i} \frac{\tilde{t}_{,i}\tilde{\rho}_{,i}}{g_{i}} + b(\theta^{i}) \left( \tilde{\rho}^{2} + \sum_{i} \frac{\tilde{\rho}_{,i}^{2}}{g_{i}} \right)}.
\end{equation}
By invoking regularity of the HRT surface at the tip, we find $a$ and $b$ to be
\begin{align}
	a{(\theta^{i})} &= 0,\\
	b{(\theta^{i})} &= \left( \frac{L}{z_m} \right)^{d-1} \tilde{\rho}^{d-2} \sqrt{\mathrm{det}{(d\Omega_{d-2}^{2})}},
\end{align}
and we find that $\rho_{0}'$ satisfies the algebraic equation
\begin{equation}
	\frac{\gamma \rho_{0}'}{\sqrt{\alpha (\rho_{0}')^{2} + \beta}} = \left( \frac{z}{z_m} \right)^{d-1},
\end{equation}
with
\begin{equation}
	\alpha = 1 - \frac{f}{\tilde{\rho}^{2}} \frac{\sum_{i} \frac{\tilde{\rho}_{,i}^{2}}{g_i} \sum_{j} \frac{\tilde{t}_{,j}^{2}}{g_j} -\left( \sum_{i} \frac{\tilde{\rho}_{,i} \tilde{t}_{,i}}{g_i} \right)^{2} + \tilde{\rho}^{2} \sum_{i} \frac{\tilde{t}_{,i}^{2}}{g_i}}{\sum_{i} \frac{\tilde{\rho}_{,i}^{2}}{g_i} + \tilde{\rho}^{2}},
\end{equation}
\begin{equation}
	\beta = \frac{1}{\tilde{\rho}^{2}} \frac{\tilde{\rho}^{2} + \sum_{i} \frac{\tilde{\rho}_{,i}^{2}}{g_i}}{f} - \frac{\sum_{i < j} \frac{(\tilde{\rho}_{,i} \tilde{t}_{,j} - \tilde{t}_{,i} \tilde{\rho}_{,j})^2}{g_{i}g_{j}} + \tilde{\rho}^{2} \sum_{i} \frac{\tilde{t}_{,i}^{2}}{g_i}}{\tilde{\rho}^{4}},
\end{equation}
\begin{equation}
	\gamma = 1 - \frac{f}{\tilde{\rho}^{2}} \sum_{i} \frac{\tilde{t}_{,i}^{2}}{g_i} + \frac{ \left( \sum_{i} \frac{\tilde{\rho}_{,i} \tilde{t}_{,i}}{g_i} \right)^{2} f }{\tilde{\rho}^{2} (\sum_{i} \frac{\tilde{\rho}_{,i}^{2}}{g_i} + \tilde{\rho}^{2})}.
\end{equation}
Solving for $\rho_{0}'$ yields
\begin{equation}
	\rho_{0}' = \frac{\sqrt{\beta} z^{d-1}}{\sqrt{z_{m}^{2d-2}\gamma^{2} - \alpha z^{2d-2}}}.
\end{equation}
Expanding the above around $z=0$,
\begin{equation}
	\rho_{0}' = \frac{\sqrt{1 +  \frac{1}{\tilde{\rho}^{2}} \sum_{i} \frac{(\tilde{\rho}_{,i}^{2} - \tilde{t}_{,i}^{2})}{g_i} - \frac{1}{\tilde{\rho}^{4}} \sum_{i<j} \frac{(\tilde{\rho}_{,i}\tilde{t}_{,j} - \tilde{t}_{,i}\tilde{\rho}_{,j})^2}{g_{i}g_{j}}}}{\left[ 1 - \frac{1}{\tilde{\rho}^{2}} \sum_{i} \frac{\tilde{t}_{,i}^{2}}{g_i} + \frac{1}{\tilde{\rho}^{2} (\sum_{i} \frac{\tilde{\rho}_{,i}^{2}}{g_i} + \tilde{\rho}^{2})} \left( \sum_{i} \frac{\tilde{\rho}_{,i}\tilde{t}_{,i}}{g_i} \right)^{2} \right]} \left( \frac{z}{z_m} \right)^{d-1} + \dots~.
\end{equation}
From the above, we see that the near-boundary expansion for $\rho_{0}'$ indeed starts with the $(d-1)$th power, as previously claimed. $z_{m}$ is near the horizon, which we will set to $1$ again. Integrating over $z$, we obtain the coefficient $c_{d}$,
\begin{equation}
	c_{d} = -\frac{1}{d} \frac{\sqrt{1 +  \frac{1}{\tilde{\rho}^{2}} \sum_{i} \frac{(\tilde{\rho}_{,i}^{2} - \tilde{t}_{,i}^{2})}{g_i} - \frac{1}{\tilde{\rho}^{4}} \sum_{i<j} \frac{(\tilde{\rho}_{,i}\tilde{t}_{,j} - \tilde{t}_{,i}\tilde{\rho}_{,j})^2}{g_{i}g_{j}}}}{\left[ 1 - \frac{1}{\tilde{\rho}^{2}} \sum_{i} \frac{\tilde{t}_{,i}^{2}}{g_i} + \frac{1}{\tilde{\rho}^{2} (\sum_{i} \frac{\tilde{\rho}_{,i}^{2}}{g_i} + \tilde{\rho}^{2})} \left( \sum_{i} \frac{\tilde{\rho}_{,i}\tilde{t}_{,i}}{g_i} \right)^{2} \right]}
\end{equation}
and also the coefficient $d_{d}$,
\begin{equation}
	d_{d} = \frac{\sum_{i} \frac{\tilde{\rho}_{,i} \tilde{t}_{,i}}{g_i}}{\tilde{\rho}^{2} + \sum_{i} \frac{\tilde{\rho}_{,i}^{2}}{g_i}} c_{d}.
\end{equation}
Plugging the coefficients above into the conjugate momenta $\Pi^{z}_{\rho}$ and $\Pi^{z}_{t}$, we find that they simplify to
\begin{align}
	\Pi^{z}_{\rho} &= -L^{d-1} R^{d-2} \tilde{\rho}^{d-2} \sqrt{\mathrm{det}{(d\Omega_{d-2}^{2})}},\\
	\Pi^{z}_{t} &= 0.
\end{align}
Plugging the above into $\frac{dA}{dR}$, we find
\begin{equation}
	\frac{dA}{dR} = L^{d-1} R^{d-2} \int d\Omega_{d-2} \sqrt{\mathrm{det}{(d\Omega_{d-2}^{2})}} \tilde{\rho}{(\theta^{i})}^{d-1}. 
\end{equation}
Let us now compare with the projected volume. This latter is
\begin{eqnarray}
	\mathrm{Volume} &=& \int d\Omega_{d-2} \sqrt{\mathrm{det}{(d\Omega_{d-2}^{2})}} \int_{0}^{R\tilde{\rho}} \rho^{d-2} d\rho \nonumber \\
	&=& \int d\Omega_{d-2} \sqrt{\mathrm{det}{(d\Omega_{d-2}^{2})}} \frac{(R\tilde{\rho})^{d-1}}{d-1}.
\end{eqnarray}
Differentiating with respect to $R$ yields
\begin{equation}
	\frac{d \mathrm{Volume}}{dR} = R^{d-2} \int d\Omega_{d-2} \sqrt{\mathrm{det}{(d\Omega_{d-2}^{2})}} \tilde{\rho}^{d-1} 
\end{equation}
which is the same as $\frac{d\mathcal{A}}{dR}$ up to the factor of entropy density $L^{d-1}$. Thus, we have managed to verify the volume-law entropy in arbitrary dimensions.

\end{document}